\documentclass[a4paper, onecolumn, preprintnumbers, nofootinbib, notitlepage]{article}
\usepackage{textcase}
\usepackage{titlesec}
\usepackage{jcappub}

\usepackage[a4paper, hdivide={2.245cm,,2.245cm}, vdivide={1.83cm,,3.8cm}]{geometry}

\usepackage{color}
\usepackage{ulem}
\usepackage{youngtab}
\usepackage[utf8]{inputenc}
\usepackage[english]{babel} 
\usepackage{amsmath}
\usepackage{float}
\usepackage{amssymb, url}

\usepackage{braket}
\usepackage{amsfonts,bbm}
\usepackage{dsfont,tensor}
\usepackage{enumerate}
\makeatletter\AtBeginDocument{\let\@elt\relax}\makeatother
\usepackage{graphicx}
\usepackage{bbm}
\usepackage{array}
\usepackage{booktabs}
\usepackage{units}
\usepackage{textcomp}
\usepackage{mathtools}
\usepackage{cancel}
\usepackage{slashed}
\usepackage{multirow}
\usepackage{slashed}
\usepackage{comment}
\usepackage{relsize}
\usepackage{tcolorbox}
\usepackage{csquotes}
\usepackage{simpler-wick}
\usepackage{lipsum}
\usepackage[compat=1.0.0]{tikz-feynman}

\allowdisplaybreaks

\newcommand{\overbar}[1]{\mkern 1.5mu\overline{\mkern-3.5mu#1\mkern-1.5mu}\mkern 2.5mu}
\newcommand{\Nbar}{\overbar{N}}

\newcommand{\0}[1]{\color{black}{#1}\color{black}}
\newcommand{\2}[1]{\color{black}{#1}\color{black}}
\newcommand{\n}[1]{\color{black}{#1}\color{black}}

\title{Matter-antimatter asymmetry and dark matter stability from baryon number conservation}

\author{Mar Císcar-Monsalvatje,}
\author{Alejandro Ibarra}
\author{and J\'{e}r\^{o}me~Vandecasteele}

\affiliation{TUM School of Natural Sciences, Technische Universität München, \\
James-Franck-Straße, 85748 Garching, Germany}

\emailAdd{mar.ciscar@tum.de}
\emailAdd{ibarra@tum.de}
\emailAdd{jerome.vandecasteele@tum.de}

\abstract{There is currently no evidence for a baryon asymmetry in our universe. Instead, cosmological observations have only demonstrated the existence of a quark-antiquark asymmetry, which does not necessarily imply a baryon asymmetric Universe, since the baryon number of the dark sector particles is unknown. 
In this paper we discuss a framework where the total baryon number of the Universe is equal to zero, and where the observed quark-antiquark asymmetry arises from neutron portal interactions with a dark sector fermion $N$ that carries baryon number. In order to render a baryon symmetric universe throughout the whole cosmological history, we introduce a complex scalar $\chi$, with opposite baryon number and with the same initial abundance as $N$.
Notably, due to the baryon number conservation, $\chi$ is absolutely stable and could have an abundance today equal to the observed dark matter abundance. Therefore, in this simple framework, the existence of a quark-antiquark asymmetry is intimately related to the existence (and the stability) of dark matter.}
\arxivnumber{2307.02592}
\begin{document}
\maketitle

\section{Introduction}

The Standard Model (SM) of Particle Physics describes with outstanding precision the results of a myriad of experiments involving particle reactions. However, several cosmological observations suggest that the SM should be extended.  Two of the most solid evidences for New Physics beyond the SM are the existence of dark matter in our Universe~\cite{Planck:2018vyg}
\begin{align}
\Omega_{\rm DM,0}h^2=0.120\pm 0.001\;,
\end{align}
and the existence of a cosmic asymmetry between the number of SM matter particles and their antiparticles, commonly expressed as the difference between the number density of baryons and antibaryons normalized to the entropy density~\cite{Planck:2018vyg}
\begin{align}
Y_{B,0}=\frac{n_B-n_{\bar B}}{s}\Big|_0=(8.75\pm 0.23)\times 10^{-11}\;.
\label{eq:Y_B0}
\end{align}
Furthermore, observations have revealed that the density in the form of dark matter is comparable to the density of Standard Model Matter (SMM) $\Omega_{\rm DM}/\Omega_{\rm SMM}\sim 5$.

In 1967, Sakharov presented three necessary conditions that must be simultaneously fulfilled in order to generate a baryon asymmetry in our Universe~\cite{Sakharov:1967dj}:  (i) baryon number violation; (ii) $C$ and $CP$ violation; (iii) departure from thermal equilibrium. Many concrete models fulfilling these three conditions have been proposed which generate a baryon asymmetry in qualitative agreement with observations (see {\it e.g.} \cite{Yoshimura:1978ex,Kuzmin:1985mm,Dodelson:1990, Farrar:1993sp, Aitken:2017wie}). On the other hand, in these models the dark matter is typically not accounted for and assumed to be produced through a different mechanism, involving different particles and interactions, and occurring at different cosmic times. Hence, the similarity between the densities of protons and dark matter is merely coincidental. A popular framework to explain this similarity consists in postulating that the dark sector is asymmetric under a global dark symmetry, $U(1)_D$, while the visible sector is asymmetric under a global baryon symmetry, $U(1)_B$. If appropriate conditions are fulfilled in the dark sector, analogous to the Sakharov conditions, an asymmetry between dark matter particle and antiparticle could be generated, which is then transferred to  the visible sector  \cite{Kaplan:1991ah,Kuzmin:1996he,Barr:1990ca,Kitano:2004sv,Bernal:2016gfn, Bringmann:2018sbs, Servant:2013uwa}. Alternatively, both asymmetries could be generated simultaneously in the dark and the visible sectors \cite{Oaknin:2005,Farrar:2006, Davoudiasl:2010am, Elor:2018twp}. 
For reviews on asymmetric dark matter, see \cite{ Davoudiasl:2012uw, Petraki:2013wwa, Zurek:2013wia}. 

From the observational standpoint one cannot conclude that the Universe is baryon asymmetric, since the baryon number of the dark sector particles is unknown. Instead, one can only asseverate that the visible sector is baryon asymmetric, or more strictly, that the Universe contains an asymmetry between the total number of quarks and antiquarks, given by
\begin{align}
Y_{\Delta q,0}= (2.63\pm 0.07)\times 10^{-10},
\end{align}
which is obtained from multiplying Eq.~(\ref{eq:Y_B0}) by 3.

In this paper we will argue that not all the Sakharov conditions are necessary to generate a quark-antiquark asymmetry. We will assume that the baryon and lepton numbers are exact symmetries of Nature, and we will show that a cosmic quark-antiquark asymmetry can be generated when the three following conditions are satisfied: (i) $C$ and $CP$ is violated in the dark sector, (ii) there is departure from thermal equilibrium, (iii) there are portal interactions between the dark sector and the quarks. These conditions are (arguably) less restrictive that the Sakharov conditions, and seem plausible in dark sector scenarios. 

To illustrate the idea, we will consider a simple framework where the dark matter particle is a complex scalar carrying baryon number, and that the dark sector interacts with the visible sector via a ``neutron portal" \cite{Fornal:2018eol}. Imposing an initial asymmetry between the number of dark matter particles and antiparticles, we will show that the asymmetry in the dark sector is transmitted to the visible sector thus generating an asymmetry between the number of quarks and antiquarks. In this way, the yield of baryons in the visible sector and the yield of dark matter particles are naturally comparable. Furthermore, in this simple framework the dark matter stability is ensured by the conservation of the baryon number, and does not require additional {\it ad hoc} symmetries. Therefore, in this scenario the existence of dark matter in our Universe today is intimately related to the existence of a quark-antiquark asymmetry.

The paper is organized as follows. In Section \ref{sec:model} we present our scenario and we qualitatively describe its main characteristics. In Section \ref{sec:Yevolve} we present the Boltzmann equations for the temperature evolution of the yields of the various particle species, and we estimate the present value of the dark matter abundance and the quark-antiquark asymmetries in terms of the initial conditions of our scenario. In Section \ref{sec:constraintNP} we discuss the constraints on the neutron portal and the prospects of detecting signals from the dark sector, and in Section \ref{sec:DMconstr} the prospects of detecting a dark matter signal. Finally, in Section \ref{sec:conclusions} we present our conclusions. 

\section{Dark sector baryons and their impact on the visible sector}
\label{sec:model}

We consider a hidden sector containing a complex scalar $\chi$ and a Dirac fermion $N$, both singlets under the Standard Model gauge group, with masses $m_\chi$ and $m_N$ respectively. We assume that these fields transform under an exactly conserved $U(1)_B$ symmetry, with charges $B(\chi)=-1$ and $B(N)=+1$. The baryon numbers of the proton and the neutron are defined as usual as $B(p)=+1$, $B(n)=+1$ and so are the baryon numbers of the quarks and antiquarks, $B(q)=+1/3$, $B(\bar q)=-1/3$.

The kinetic and mass terms in the Lagrangian of the dark sector baryons read
\begin{align}\label{eq:LagKin}
\mathcal{L}\supset \left|\partial_\mu\chi\right|^2+m_\chi^2 \chi^*\chi + \Nbar\left(i\slashed{\partial}-m_N\right)N.
\end{align}
Further, the Lagrangian contains interaction terms among the dark sector particles, which we describe via dimension-5 effective operators of the form 
\begin{align}\label{eq:LagInter}
\mathcal{L}\supset  \frac{1}{\0{\Lambda_0}}\chi\chi^*\Nbar N+\frac{1}{\2{\Lambda_2}}\left(\chi \chi \overbar{N^c}N +\, h.c.\right),
\end{align}
where the superscript $c$ denotes charge conjugation.
The first term does not change the baryon number in the fermionic current, while the second term changes the baryon number by two units, hence the notation for the suppression scale of the dimension-5 operators,  $\0 {\Lambda_0}$ and $\2{\Lambda_2}$ respectively.

Lastly, the gauge symmetry and the baryon symmetry allow interaction terms between the dark sector and the visible sector of the form 
\begin{align}\label{eq:LagNeutronPortal}
\mathcal{L}\supset \lambda_{\chi H}|\chi|^2 |H|^2+ \frac{1}{\Lambda_n^2}\left(\Nbar d_R \, \overline{u_R^c} d_R+\, h.c\right),
\end{align}
with $H$ the Standard Model Higgs doublet, and $u_R$ and $d_R$ the right-handed up and down quarks. These two terms respectively correspond to a Higgs portal interaction and to a neutron portal interaction \cite{Davoudiasl:2010am, Fornal:2018eol, Fornal:2023wji}. Portals involving heavier generation quarks are also possible, but will not be discussed in what follows for simplicity.

\begin{figure}
\centering
\includegraphics[width=0.5\textwidth]{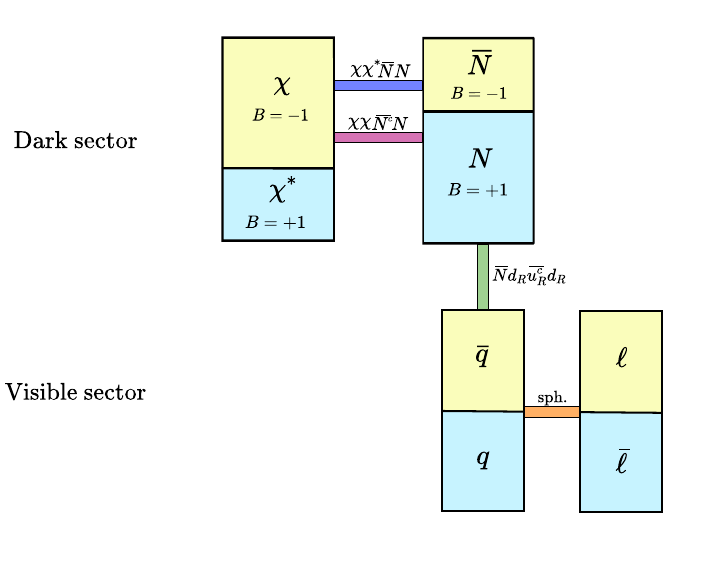}
\caption{Sketch of the initial abundances of the different particle species in our model, together with the interaction terms among them. Light yellow (blue) indicates the particles with $B-L<0$ ($B-L>0$); the total $B-L$ of the Universe is equal to zero. 
}\label{fig:boxdiag_initial}
\end{figure}

 In order to generate an asymmetry in the visible sector, we assume that some unspecified mechanism in the dark sector produces an excess of $N$ over $\overbar N$ at high temperatures. Since this mechanism operates exclusively in the dark sector, the conservation of baryon number requires the generation of an excess of $\chi$ over $\chi^*$ so that the total baryon number of the Universe remains zero, as depicted in Fig.~\ref{fig:boxdiag_initial}. A possible mechanism generating this excess could be the out-of-equilibrium, $CP$-violating, and $B$-conserving decays of heavy fermions $\varphi_i\rightarrow \chi N, \chi^* \overbar{N}$, which will generate an asymmetry between the number densities of $N$ and $\overbar{N}$, and the corresponding asymmetry between $\chi$ and $\chi^*$, rendering a total baryon number equal to zero. A quantitative description of this mechanism will be presented elsewhere \cite{future}. 
 
 Due to the neutron portal interaction in Eq.~(\ref{eq:LagNeutronPortal}), the scatterings $N \bar d \leftrightarrow u d$ and $N \bar u \leftrightarrow d d$, as well as the decays $N\rightarrow u d d$, inject a net baryon number into the visible sector (which could be partially converted into a net lepton number via sphaleron transitions,  depending on the temperature at which the asymmetry in the dark sector is generated). Therefore, the excess of dark sector fermionic baryons over antibaryons is leaked to the visible sector via the neutron portal, ultimately generating an excess of quarks over antiquarks.

Note that the Higgs portal does not transmit baryon number from the dark sector to the visible sector. Hence, we will set it to zero for simplicity, although it could have phenomenological implications, as we will briefly discuss in Section \ref {sec:DMconstr}.

Notably, due to the Lorentz symmetry and the baryon number conservation, $\chi$ and $\chi^*$ are absolutely stable, although they can annihilate with one another generating a relic population of $\chi$. Therefore, in this simple framework the existence of a quark-antiquark asymmetry in the visible sector is intimately related to the existence of dark matter in our Universe. We stress that the stability of the dark matter does not require any {\it ad hoc} new symmetry, but is simply due to the conservation of the total baryon number in the Universe.\footnote{In the model presented in Eq.~(\ref{eq:LagKin}$-$\ref{eq:LagNeutronPortal}), the dark matter particle exhibits an accidental $\mathbb{Z}_2$ symmetry, $\chi \rightarrow -\chi$, which however has no influence on the dark matter stability. As it has no importance for the phenomenology at hand, we will not address it further in the remaining of this work.}

In the next section, we will describe in detail the evolution of the yields of the different particles, and the expectations for the relic abundance of dark matter and the quark-antiquark asymmetry. 
\vspace{-0.2cm}

\section{Evolution of the particle number densities and asymmetries}\label{sec:Yevolve}

The Boltzmann equations for the yields of the different particle species can be cast as 
\begin{align}
 \frac{d Y_{\chi}}{dx} = - \frac{\lambda}{x^2} \Bigg[&   \0{\langle \sigma v \rangle _{\chi\chi^\ast\to N \overbar{N}}}  \left(Y_{\chi} Y_{\chi^\ast} - \frac{Y_{\chi}^{\rm eq}Y_{\chi^\ast} ^{\rm eq}}{Y_{N}^{\rm eq}Y_{\overbar{N}} ^{\rm eq}}Y_{N}Y_{\overbar{N}}\right)+2\2{\langle \sigma v \rangle _{\chi\chi\to \overbar{N} \overbar{N}}} \left(Y_{\chi}^2 - \left(\frac{Y_{\chi}^{\rm eq}}{Y_{\overbar{N}}^{\rm eq}}\right)^{2}Y_{\overbar{N}}^2\right) \nonumber \\[5pt]
 &+\2{\langle \sigma v \rangle _{\chi N\to \chi^\ast\overbar{N}}} \Bigg(Y_{\chi}Y_{N} - \frac{Y_{\chi}^{\rm eq}Y_{N} ^{\rm eq}}{Y_{\chi^\ast} ^{\rm eq}Y_{\overbar{N}} ^{\rm eq}}Y_{\chi^\ast}Y_{\overbar{N}}\Bigg)\Bigg],\\[10pt]
\label{eq:Yk}
 \frac{d Y_{\chi^\ast}}{dx} = - \frac{\lambda}{x^2} \Bigg[ &  \0{\langle \sigma v \rangle _{\chi\chi^\ast\to N \overbar{N}}}\left(Y_{\chi} Y_{\chi^\ast} - \frac{Y_{\chi}^{\rm eq}Y_{\chi^\ast} ^{\rm eq}}{Y_{N}^{\rm eq}Y_{\overbar{N}} ^{\rm eq}}Y_{N}Y_{\overbar{N}}\right)+2\2{\langle \sigma v \rangle _{\chi^\ast\chi^\ast\to N N}} \Bigg(Y_{\chi^\ast}^2 - \left(\frac{Y_{\chi^\ast}^{\rm eq}}{Y_{N}^{\rm eq}}\right)^{2}Y_{N}^2\Bigg) \nonumber \\[5pt]
 & -\2{\langle \sigma v \rangle _{\chi N\to \chi^\ast\overbar{N}}} \Bigg(Y_{\chi}Y_{N} - \frac{Y_{\chi}^{\rm eq}Y_{N} ^{\rm eq}}{Y_{\chi^\ast} ^{\rm eq}Y_{\overbar{N}} ^{\rm eq}}Y_{\chi^\ast}Y_{\overbar{N}}\Bigg)\Bigg], \\[10pt]
  \frac{d Y_{N}}{dx} = - \frac{\lambda}{x^2} \Bigg[  & -\0{\langle \sigma v \rangle _{\chi\chi^\ast\to N \overbar{N}} }\left(Y_{\chi} Y_{\chi^\ast} - \frac{Y_{\chi}^{\rm eq}Y_{\chi^\ast} ^{\rm eq}}{Y_{N}^{\rm eq}Y_{\overbar{N}} ^{\rm eq}}Y_{N}Y_{\overbar{N}}\right)-2\2{\langle \sigma v \rangle _{\chi^\ast\chi^\ast\to N N}} \left(Y_{\chi^\ast}^2 - \left(\frac{Y_{\chi^\ast}^{\rm eq}}{Y_{N}^{\rm eq}}\right)^{2}Y_{N}^2\right) \nonumber \\[5pt]
  &+\2{\langle \sigma v \rangle _{\chi N\to \chi^\ast\overbar{N}}} \left(Y_{\chi}Y_{N} - \frac{Y_{\chi}^{\rm eq}Y_{N} ^{\rm eq}}{Y_{\chi^\ast} ^{\rm eq}Y_{\overbar{N}} ^{\rm eq}}Y_{\chi^\ast}Y_{\overbar{N}}\right) +\frac{1}{\mathfrak{s}}\n{\langle\Gamma_N\rangle}\Big(Y_N - Y_N^{\rm eq}\Big) \nonumber
  \\[5pt]
  &+\n{\langle \sigma v \rangle _{N\bar{d}\to ud}}Y_{\bar{d}}^{\rm eq}\Big(Y_{N} -Y_{N}^{\rm eq}  \Big)
  +\n{\langle \sigma v \rangle _{N\bar{u}\to dd}}Y_{\bar{u}}^{\rm eq}\Big(Y_{N} -Y_{N}^{\rm eq} \Big)\Bigg],\\[10pt]
\frac{d Y_{\overbar{N}}}{dx} = - \frac{\lambda}{x^2} \Bigg[&  -\0{\langle \sigma v \rangle _{\chi\chi^\ast\to N \overbar{N}}}\left(Y_{\chi} Y_{\chi^\ast} - \frac{Y_{\chi}^{\rm eq}Y_{\chi^\ast} ^{\rm eq}}{Y_{N}^{\rm eq}Y_{\overbar{N}} ^{\rm eq}}Y_{N}Y_{\overbar{N}}\right)-2\2{\langle \sigma v \rangle _{\chi\chi\to \overbar{N} \overbar{N}}} \left(Y_{\chi}^2 - \left(\frac{Y_{\chi}^{\rm eq}}{Y_{\overbar{N}}^{\rm eq}}\right)^{2}Y_{\overbar{N}}^2\right) \nonumber \\[5pt]
&-\2{\langle \sigma v \rangle _{\chi N\to \chi^\ast\overbar{N}}} \left(Y_{\chi}Y_{N} - \frac{Y_{\chi}^{\rm eq}Y_{N} ^{\rm eq}}{Y_{\chi^\ast} ^{\rm eq}Y_{\overbar{N}} ^{\rm eq}}Y_{\chi^\ast}Y_{\overbar{N}}\right)+\frac{1}{\mathfrak{s}}\n{\langle\Gamma_N\rangle}\Big(Y_{\overbar{N}} - Y_{\overbar{N}}^{\rm eq}\Big) \nonumber\\[5pt]
&+\n{\langle \sigma v \rangle _{\overbar{N}d\to \bar u\bar d}}Y_{d}^{\rm eq}\Big(Y_{\overbar{N}}-Y_{\overbar{N}}^{\rm eq} \Big)
+\n{\langle \sigma v \rangle _{\overbar N u\to \bar d \bar d}}Y_{u}^{\rm eq}\Big(Y_{\overbar{N}}-Y_{\overbar{N}}^{\rm eq} \Big)\Bigg] \;.
\end{align}
Here $x\equiv m_{\chi}/T$, $\mathfrak{s}=2\pi^2/45 \,g_{\star, \mathfrak{s}}T^3$ is the entropy density of the Universe at the temperature $T$, with $g_{\star, \mathfrak{s}}$ the number of relativistic degrees of freedom. The overall numerical factor $\lambda$ is defined as 
\begin{equation}\label{eq:lambda}
   \lambda \equiv \frac{4 \pi }{\sqrt{90} }~m_{\chi} M_{\rm Pl}~\sqrt{g_{\star, \mathfrak{s}}}\,.
\end{equation}

The Boltzmann equations depend on the thermally averaged rates of different processes. Firstly, there are reactions involving only dark sector particles: $\chi\chi^*\leftrightarrow N\overbar N$, which is induced by the effective operator in the first term of Eq.~(\ref{eq:LagInter}), suppressed by $\0{\Lambda_0}$;  $\chi\chi\leftrightarrow \overbar N\overbar N$ and the $C-$conjugated reaction  $\chi^*\chi^* \leftrightarrow N N$, induced by the effective operator found in the second term of Eq.~(\ref{eq:LagInter}), suppressed by $\2{\Lambda_2}$; and $\chi N \leftrightarrow \chi^* \overbar N$, also induced by the same term and suppressed by $\2{\Lambda_2}$. The explicit expressions for the cross sections of these processes read
\begin{align}
\0{\sigma _{\chi\chi^\ast\to N \overbar{N}}}&=  \frac{\left(s-4 m_N^2  \right)^{3/2}}{8\pi\, s \sqrt{s-4 m_\chi ^2}\,
\Lambda_0^2}, \label{eq:chichi*}\\
    \2{\sigma  _{\chi\chi\to \overbar{N} \overbar{N}}}&=\2{ \sigma  _{\chi^\ast\chi^\ast\to N N}}=  \frac{\left(s-4 m_N^2  \right)^{3/2}}{8\pi\, s \sqrt{s-4 m_\chi ^2}\, \Lambda_{2}^2},\label{eq:chichi}\\
 \2{\sigma  _{\chi N\to \chi^\ast\overbar{N}}}&=\frac{m_N^4-2 m_N^2 \left(m_\chi^2 -3 s \right) + \left(m_\chi^2 - s\right)^2}{32\pi\, s^2 \Lambda_2^2},
\label{eq:chiN}
\end{align}
with $s$ the square of the center of mass energy.  Secondly, one has reactions between dark sector particles and visible sector particles,  $N \bar{d}\leftrightarrow ud$ and  $N \bar{u}\leftrightarrow dd$, as well as the decay $N\rightarrow u d d$, all mediated by the neutron portal interaction, suppressed by $\Lambda_n$. The corresponding cross sections and decay rate read
\begin{align}
\n{\sigma_{ N \bar{d}\rightarrow ud}}&=\frac{m_N^2+14s}{192\pi \Lambda_n^4}\label{eq:sigmaNeutronPortal}, \\
\n{\sigma_{ N \bar{u}\rightarrow dd}}&=\frac{2m_N^2+s}{192\pi \Lambda_n^4},\\
\n{\Gamma_N}&=\frac{1}{768\pi^3}\frac{m_N^5}{\Lambda_n^4}.
\end{align}

To simplify the discussion, we will assume that the neutron portal interaction is sufficiently strong to bring the dark sector baryons into thermal equilibrium with the visible sector. At a temperature $T$, this condition requires
\begin{align}
\n{\Lambda_n}\lesssim 5.8\color{black}\times 10^7 \left(\frac{T}{10^5\text{ GeV}}\right)^{3/4} \text{ GeV}.
\label{eq:thermalization}
\end{align}
Therefore the equilibrium number density of the dark sector particle species $i$ with mass $m_i$, number of internal degrees of freedom $g_i$ and chemical potential $\mu_i$ is given as usual by
\begin{align}
    n_i^{\text{eq}}=\frac{g_i}{2\pi^2}m_i^2\,T K_2\left(m_i/T\right)\mathrm{e}^{\mu_i/T},\label{eq:neq}
\end{align}
where $K_2(x)$ is the modified Bessel function of the $2^{\text{nd}}$ kind. 

It will be convenient to work with the total yields of the different particle species, along with their corresponding asymmetries, defined as
\begin{align}
&Y_{\chi}^{\text{tot}}=Y_\chi+Y_{\chi^\ast },
&&Y_{\Delta\chi}=Y_\chi-Y_{\chi^\ast}, \\
&Y_{N}^{\text{tot}}=Y_N+Y_{\Nbar},
&&Y_{\Delta N}=Y_N-Y_{\Nbar}.
\end{align}
The  Boltzmann equations describing their temperature evolution are
\begin{align}
& \frac{d Y_{\chi}^{\rm tot}}{dx} =  - \frac{\lambda}{x^2} \Bigg[  2\0{\langle \sigma v \rangle _{\chi\chi^\ast\to N \overbar{N}}}\Big(Y_{\chi} Y_{\chi^\ast} - P Y_{N}Y_{\overbar{N}}\Big)+2\2{\langle \sigma v \rangle _{\chi\chi\to \overbar{N} \overbar{N}}} \Big(Y_{\chi}^2 +Y_{\chi^\ast}^2 - P \left(Y_N^2+Y_{\overbar{N}}^2\right)\Big)\Bigg],\label{eq:dYdchitot}\\[12pt]
& \frac{d Y_{\Delta\chi}}{dx} =- \frac{\lambda}{x^2} \Bigg[   2\2{\langle \sigma v \rangle _{\chi\chi\to \overbar{N} \overbar{N}}}\Big(Y_{\chi}^2 -Y_{\chi^\ast}^2 + P \left(Y_N^2-Y_{\overbar{N}}^2\right)\Big)  +2\2{\langle \sigma v \rangle _{\chi N\to \chi^\ast\overbar{N}}} \Big(Y_{\chi}Y_{N}- Y_{\chi^\ast}Y_{\overbar{N}}\Big)\Bigg],\label{eq:dYdeltachidx} \\[12pt]
 & \frac{d Y_{N}^{\rm tot}}{dx} = - \frac{\lambda}{x^2} \Bigg[   -2\0{\langle \sigma v \rangle _{\chi\chi^\ast\to N \overbar{N}}}\Big(Y_{\chi} Y_{\chi^\ast} - P Y_{N}Y_{\overbar{N}}\Big)-2\2{\langle \sigma v \rangle _{\chi\chi\to \overbar{N} \overbar{N}}}\Big(Y_{\chi}^2 +Y_{\chi^\ast}^2 - P \left(Y_N^2+Y_{\overbar{N}}^2\right)\Big) \label{eq:dYNtot}\\
  &+\frac{1}{\mathfrak{s}}\n{\langle\Gamma_N\rangle}\Big(Y_{N}  +Y_{\overbar{N}}- Y_{N}^{\rm eq}- Y_{\overbar{N}}^{\rm eq}\Big)+\n{\langle \sigma v \rangle _{N\bar{d}\to ud}}\Big(Y_{N} Y_{\bar{d}}^{\rm eq}+Y_{\overbar{N}}Y_{d}^{\rm eq} -Y_{N}^{\rm eq}  Y_{\bar{d}}^{\rm eq} -Y_{\overbar{N}}^{\rm eq} Y_{d}^{\rm eq} \Big)+(N\bar u \leftrightarrow dd)\Bigg],\nonumber\\[12pt]
&\frac{d Y_{\Delta N}}{dx} =  - \frac{\lambda}{x^2} \Bigg[  2\2{\langle \sigma v \rangle _{\chi\chi\to \overbar{N} \overbar{N}}}\Big(Y_{\chi}^2 -Y_{\chi^\ast}^2 + P \left(Y_N^2-Y_{\overbar{N}}^2\right)\Big)+2\2{\langle \sigma v \rangle _{\chi N\to \chi^\ast\overbar{N}}} \Big(Y_{\chi}Y_{N} -Y_{\chi^\ast}Y_{\overbar{N}}\Big)\label{eq:dYDeltaNdx}\\
&+\frac{1}{\mathfrak{s}}\n{\langle \Gamma_N\rangle}\Big(Y_{N}  -Y_{\overbar{N}}- Y_{N}^{\rm eq}+ Y_{\overbar{N}}^{\rm eq}\Big)+\n{\langle \sigma v \rangle _{N\bar{d}\to ud}}\Big(Y_{N} Y_{\bar{d}}^{\rm eq}-Y_{\overbar{N}}Y_{d}^{\rm eq} -Y_{N}^{\rm eq} Y_{\bar{d}}^{\rm eq} + Y_{\overbar{N}}^{\rm eq}Y_{d}^{\rm eq} \Big)+(N\bar u \leftrightarrow dd)\Bigg].\nonumber
\end{align}
The yields $Y_\chi$, $Y_{\chi^\ast}$, $Y_N$ and $Y_{\Nbar}$ are implicit functions of the total yields and of the asymmetries. Further, we have introduced
\begin{align}
   P\,\equiv\, \frac{Y_{\chi}^{\rm eq}Y_{\chi^\ast} ^{\rm eq}}{Y_{N}^{\rm eq}Y_{\overbar{N}} ^{\rm eq}}=\left(\frac{Y_\chi^{\text{eq}}}{Y_{\overbar{N}}^{\text{eq}}}\right)^2=\Bigg(\frac{Y_{\chi^\ast}^{\text{eq}}}{Y_N^{\text{eq}}}\Bigg)^2
   \simeq \left(\frac{g_\chi}{g_N}\right)^2\left(\frac{m_\chi}{m_N}\right)^4\Bigg(\frac{K_2\left(m_\chi/T\right)}{K_2\left(m_N/T\right)}\Bigg)^2.
\end{align}
Lastly, the asymmetry present in $N$ is transmitted to the visible sector through the neutron portal, giving rise to an asymmetry between the total number of quarks and antiquarks $ \Delta q\equiv \sum_i \Delta q_i$. Its time evolution is described by
\begin{align}
       \frac{d Y_{\Delta q}}{dx} =&\,c\, \frac{3\lambda}{x^2}  \Bigg[ \frac{1}{\mathfrak{s}}\n{\langle\Gamma_N\rangle}\Big(Y_{N}  -Y_{\overbar{N}}- Y_{N}^{\rm eq}+ Y_{\overbar{N}}^{\rm eq}\Big)\label{eq:YB} \\
       &
       +\n{\langle \sigma v \rangle _{N\bar{d}\to ud}}\Big(Y_{N} Y_{\bar{d}}^{\rm eq}-Y_{\overbar{N}}Y_{d}^{\rm eq} -Y_{N}^{\rm eq} Y_{\bar{d}}^{\rm eq} + Y_{\overbar{N}}^{\rm eq}Y_{d}^{\rm eq} \Big) +(N\bar u \leftrightarrow dd)\nonumber \Bigg]
       .
\end{align}
The constant $c$ in this expression characterises the efficiency of the conversion of the baryon asymmetry stored in the left-handed quarks into a lepton asymmetry stored in the left-handed leptons via sphaleron processes. If the asymmetry in the dark sector is generated at temperatures above 130 GeV, the point at which sphalerons drop out of thermal equilibrium \cite{DOnofrio:2014rug}, then $c= 36/111$ \cite{Harvey:1990qw}. 
In turn, the asymmetry between the total number of leptons and antileptons, $ \Delta \ell \equiv \sum_i \Delta \ell_i$ can be calculated from
\begin{align}
        \frac{d Y_{\Delta \ell}}{dx} =
        -\frac{1}{3}\,\frac{1-c}{c} \,    \frac{d Y_{\Delta q}}{dx}.\label{eq:YL}
\end{align}
If the initial asymmetry is instead generated after sphaleron freeze-out, then we set $c=1$ and no asymmetry is transferred to the leptons.

Let us consider a scenario where both the baryon and the lepton numbers are exactly conserved in Nature.  
As initial condition, we assume that at a temperature $T_{\rm in}\gg 130$ GeV there is no asymmetry between quarks and antiquarks (nor between leptons and antileptons) but that there is a $CP$-violating mechanism in the dark sector that generates a primordial asymmetry between $N$ and $\overbar{N}$. The conservation of the baryon number requires a corresponding asymmetry between $\chi$ and $\chi^*$. This initial condition  is sketched in Fig.~\ref{fig:boxdiag_initial}, where we also show the different portals relating the various particle species in our model. Under this plausible assumption, the dark matter relic abundance and the quark-antiquark asymmetry are determined by the initial asymmetry in the dark sector, and by the energy scales $\Lambda_0$, $\Lambda_2$ and $\Lambda_n$, which determine the strengths of the different portal interactions. 

\begin{figure}[t]
    \centering
    \includegraphics{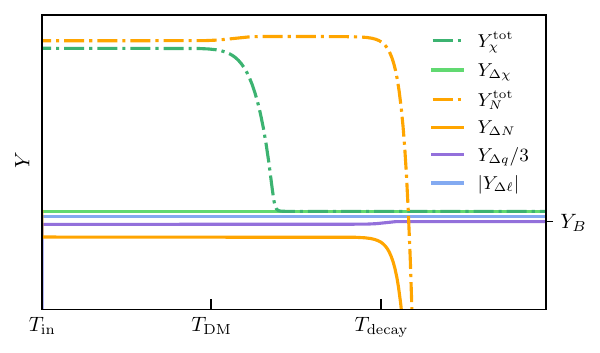}
    \caption{Representative evolution of the total yields of $\chi+\chi^*$ and  $N+\overbar{N}$ (dash-dotted) and the yields of the different asymmetries between particles and antiparticles (solid) for the initial condition depicted in Fig.~\ref{fig:boxdiag_initial}. See the main text for details.}
    \label{fig:Everything}
\end{figure}

For this representative set-up, the various yields qualitatively evolve with temperature as shown  Fig.~\ref{fig:Everything}. At the high temperature $T_{\rm in}$ all particle species are ultra-relativistic and their  equilibrium yields are related solely by their different number of internal degrees of freedom, see Eq.~(\ref{eq:neq}). For the yields of the particles in the hidden sector, this implies $Y_N^{\text{eq}}= 2 Y_\chi^{\text{eq}}$. Further, for the initial condition sketched in Fig.~\ref{fig:boxdiag_initial} the yields of the asymmetries satisfy
\begin{align}
    &Y_{\Delta N}^{\rm in}= 
    Y_{\Delta \chi}^{\rm in},\label{eq:InitialDeltaNDeltaChi} \\
    &Y_{\Delta q}^{\rm in}= 
    Y_{\Delta \ell}^{\rm in}=0.
\end{align}

Let us first discuss the evolution of the asymmetries with the temperature. At  temperatures very close to $T_{\rm in}$, the scatterings $N \bar u\leftrightarrow dd$ and $N \bar d\leftrightarrow ud $ (with rates depending on $\Lambda_n$) effectively transfer a fraction of the baryon asymmetry from the dark sector to the visible sector, which is then distributed between leptons and quarks by sphaleron transitions\footnote{In addition, fast electroweak processes share the asymmetry between all generations of quarks and leptons, thus also impacting the baryon asymmetry held in $N$ \cite{Harvey:1990qw}.}. At the same time, the processes $\chi\chi\leftrightarrow \overbar{N} \overbar{N}$ and $\chi N\leftrightarrow \chi^*\overbar{N}$ (with rates depending on $\Lambda_2$) contribute to the washout of the asymmetry within the hidden sector, thereby influencing the size of the asymmetries in the visible sector. The effect of the washout in $\Delta N$ follows from Eq.~(\ref{eq:dYDeltaNdx}). Given that at high temperatures all the yields can be well approximated by their equilibrium values, the Boltzmann equation for $Y_{\Delta N}$ simplifies to
\begin{align}  \label{eq:Y4Alone}
       \frac{d Y_{\Delta N}}{dx} \simeq -\frac{3\lambda}{x^2}\left(\2{\langle \sigma v \rangle _{\chi\chi\to \overbar{N} \overbar{N}}}+\2{\langle \sigma v \rangle _{\chi N\to \chi^\ast\overbar{N}}}\right)Y_N^{\text{eq}}Y_{\Delta N},
\end{align}
and similarly for $Y_{\Delta \chi}$. The analytical solution for this equations reads
\begin{align} \label{eq:Y4Analytical_noNP}
Y_{\Delta N}\left(T\right)\simeq 
Y_{\Delta N}^{\text{in}}\exp \Bigg\{3\bigg[\frac{n_N^{\text{eq}}}{H }\left.\left(\2{\langle \sigma v \rangle _{\chi\chi\to \overbar{N} \overbar{N}}}+\2{\langle \sigma v \rangle _{\chi N\to \chi^\ast\overbar{N}}}\right) \bigg]\right|_{T=T_{\text{in}}} \frac{T-T_{\rm in}}{T_{\text{in}}}\Bigg\}. 
\end{align}
The requirement that no more than 50\% of the asymmetry is washed out implies the lower limit\footnote{This value does not significantly change if one takes into consideration the redistribution of the asymmetry held in $N$ into the quark sector due to the neutron portal and into the lepton sector due to sphaleron conversions.}
\begin{align}
\2{\Lambda_2}\gtrsim 6\times 10^{10}\left(\frac{T_{\text{in}}}{10^5\text{ GeV}}\right)^{1/2}\text{ GeV}.\label{eq:L2wo} 
\end{align}

In order to simplify our discussion, we will assume in what follows that $\2{\Lambda_2}$ satisfies this lower limit and that the initial asymmetry in $\Delta N$ (and $\Delta \chi$) is very weakly washed out. The efficient redistribution of the asymmetry within the visible sector has been displayed in Fig.~\ref{fig:Everything}.

If that is the case, the asymmetries in the different particles species at a temperature slightly below $T_{\rm in}$ take the simple expressions
\begin{align}
    &Y_{\Delta N}(T)=\frac{11}{122} Y_{\Delta N}^{\rm in},\label{eq:NNin}\\
    &Y_{\Delta q}(T)=\frac{54}{61} Y_{\Delta N}^{\rm in},\label{eq:qNin}\\
   & Y_{\Delta \ell}(T)=-\frac{75}{122} Y_{\Delta N}^{\rm in}.
\end{align}
These asymmetries remain constant until a temperature $T_{\rm decay}$ at which the decays of $N$ and $\overbar{N}$ inject an additional quark-antiquark asymmetry in the visible sector (and a lepton-antilepton asymmetry if the sphalerons are still in equilibrium at the epoch of decays). For the range of parameters of interest (see Section \ref{sec:constraintNP}), we find that the decays typically occur when the sphalerons are out-ot-equilibrium, so that the remaining $\Delta N$ asymmetry is entirely converted into a quark-antiquark asymmetry, whereas the lepton-antilepton asymmetry stays frozen. The decrease in $Y_{\Delta N}$ at $T\sim T_{\rm in}$ and then at $T\sim T_{\rm decay}$, along with the corresponding changes in $Y_{\Delta q}$ and $Y_{\Delta \ell}$, can be seen in Fig.~\ref{fig:Everything}. At the present time, the asymmetries in the visible sector read
\begin{align}
    Y_{\Delta q,0}&=3\frac{11}{122} Y_{\Delta N}^{\rm in}+\frac{54}{61} Y_{\Delta N}^{\rm in}=\frac{141}{122}Y_{\Delta N}^{\rm in}, \label{eq:YDeltaQ-0} \\
    Y_{\Delta\ell,0}&=-\frac{75}{122} Y_{\Delta N}^{\rm in}.\label{eq:LNin}
\end{align}
In these equations, the factor of 3 arises from the fact that $N$ generates three quarks in its decay. 

Let us now turn to the evolution of the total yields $Y_{\chi}^{\rm tot}$ and $Y_{N}^{\rm tot}$, Eqs.~(\ref{eq:dYdchitot}) and (\ref{eq:dYNtot}). Again, to simplify our discussion we will assume that $\Lambda_2$ satisfies the condition in Eq.~(\ref{eq:L2wo}). Further, we will assume $\0{\Lambda_0}\ll\n{\Lambda_n}$, so that the rate of conversion of $N$ into quarks is slow compared to the rate of conversion of $\chi$ into $N$. This implies that the dark matter abundance is determined by the freeze-out of the annihilation process $\chi\chi^*\rightarrow N\overbar{N}$. Under these assumptions the Boltzmann equations (\ref{eq:dYdchitot}) and (\ref{eq:dYNtot}) simplify to

\begin{align}
 \frac{d Y_{\chi}^{\rm tot}}{dx} =&  - \frac{\lambda}{x^2} \Bigg[  2\0{\langle \sigma v \rangle _{\chi\chi^\ast\to N \overbar{N}}}\Big(Y_{\chi} Y_{\chi^\ast} - P Y_{N}Y_{\overbar{N}}\Big)\Bigg],\label{eq:DMFOchi}\\
  \frac{d Y_{N}^{\rm tot}}{dx} =& - \frac{\lambda}{x^2} \Bigg[   -2\0{\langle \sigma v \rangle _{\chi\chi^\ast\to N \overbar{N}}}\Big(Y_{\chi} Y_{\chi^\ast} - P Y_{N}Y_{\overbar{N}}\Big)\Bigg],\label{eq:DMFON}
\end{align}
which amount to a secluded sector freeze-out scenario,\footnote{If the effective neutron portal is above $\n{\Lambda_n} \approx 4\color{black}\times 10^{3}\text{ GeV}$, $N$ conversions into quarks drop out of equilibrium before the time of dark matter freeze-out. Given the experimental constraints, see Section \ref{sec:constraintNP}, the dark sector is indeed secluded from the visible sector at the time of dark matter freeze-out.} albeit with a hidden sector temperature identical to the one of the visible sector. The behaviour of $Y_\chi^{\text{tot}}$ and $Y_N^{\text{tot}}$ is sketched in Fig.~\ref{fig:Everything}. The total dark matter abundance at the current epoch can be calculated using standard tools (see {\it e.g.} \cite{Gondolo:1990dk}), resulting in
\begin{align}
\Omega_{\rm DM,0} h^2\simeq 2.8\times 10^8  \,Y_{\rm \chi}^{\rm tot}(x_{\rm f.o.}) \frac{m_\chi}{\rm GeV},
\label{eq:OmegaDM-0}
\end{align}
where the freeze-out temperature is determined by the condition  $\Gamma_{\chi\chi^\ast\to N \overbar{N}}(x_{\rm f.o.})=H(x_{\rm f.o.})$, and  $Y_{\chi}^{\rm tot}(T_{\rm f.o.})$ can be well approximated by its equilibrium value at freeze-out. Besides, due to the fact that $\chi\chi^*$ pairs can only annihilate into $N\overbar{N}$ pairs, one finds that $Y_{N}^{\rm tot}(x_{\rm f.o.})\simeq 3Y_N^{\text{eq}}\left(T_{\text{in}}\right)$, which could additionally lead to an epoch of early matter domination if $N$ is sufficiently long-lived~\cite{Cheung:2013hza,Allahverdi:2020bys,Allahverdi:2021grt}. Some implications of this early phase of matter domination will be discussed in Section \ref{sec:constraintNP}.

\begin{figure}
\centering
\includegraphics[width=0.5\textwidth]{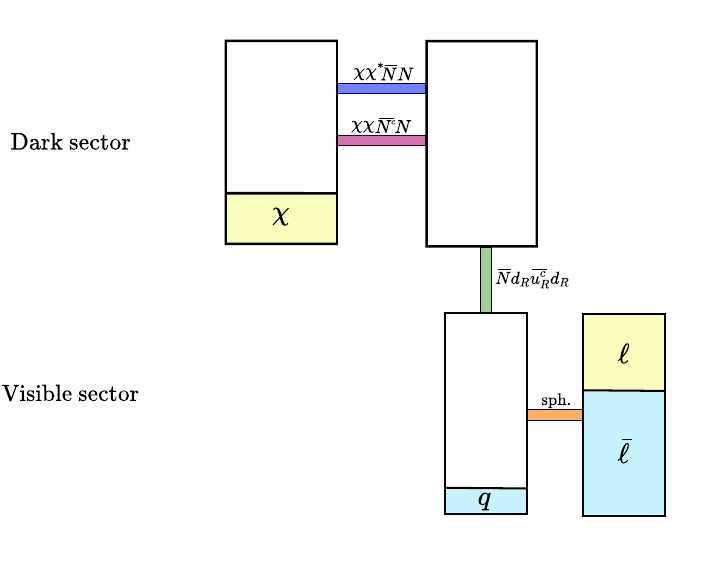}\hspace{0.5cm}
\includegraphics[width=0.345\textwidth]{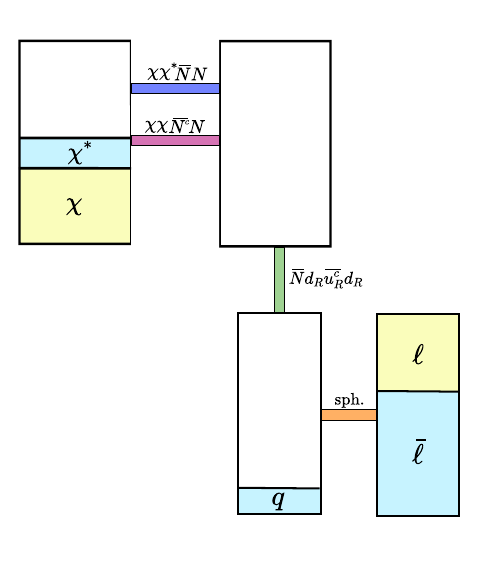}\hspace{1cm}
\caption{Possible final states from the initial state Fig.~\ref{fig:boxdiag_initial}, corresponding to a scenario where dark matter antiparticles have efficiently annihilated (left panel), or only partially annihilated (right panel). All $N$ have decayed into quarks. We have assumed that the initial asymmetry within the dark sector is generated while sphalerons are still in thermal equilibrium. Today, most leptons are in the form of neutrinos and anti-neutrinos, with a small relative asymmetry, leaked via sphaleron processes. In both cases, the total $B-L$ number of the Universe is conserved and equal to zero.
}\label{fig:boxdiag_final}
\end{figure}

In the simplest scenario, all dark matter antiparticles annihilate resulting in the final state sketched in the left diagram of  Fig.~\ref{fig:boxdiag_final}. In this case, $Y_{ \chi}^{\rm tot}(x_{\rm f.o.})=Y_{\Delta \chi}(x_{\rm f.o.})=Y_{\Delta \chi}^{\rm in}=Y_{\Delta N}^{\rm in}$
where in the last step we have used Eq.~(\ref{eq:InitialDeltaNDeltaChi}). Therefore, in this simple scenario the dark matter abundance and the quark-antiquark asymmetry are determined by the same parameter, the initial asymmetry in the dark sector, $Y_{\Delta N}^{\rm in}$. More concretely, using Eqs.~(\ref{eq:YDeltaQ-0}) and (\ref{eq:OmegaDM-0}),  one finds the relation
\begin{align} 
\Omega_{\rm DM,0}h^2\simeq 2.4\times 10^8\, Y_{\Delta q,0} \frac{m_\chi}{\rm GeV}\;.
\label{eq:relation-Omega-qbarq}
\end{align}
One can then adjust the initial condition $Y_{\Delta \chi}^{\rm in}$ to generate the observed quark-antiquark asymmetry, $Y_{\Delta q,0}\simeq 2.6\times 10^{-10}$, and the dark matter mass to generate the observed dark matter abundance, $\Omega_{\rm DM,0} h^2\simeq 0.12$. We obtain
\begin{align}
Y_{\Delta \chi}^{\rm in}&\simeq2.3\times 10^{-10},\\
m_\chi &\simeq 1.9\, \text{GeV}.\label{eq:ADMmass}
\end{align}

In the case where not all dark matter antiparticles annihilate, then $Y_{ \chi}^{\rm tot}(x_{\rm f.o.})=Y_{\Delta \chi}(x_{\rm f.o.})+2Y_{\chi^*}(x_{\rm f.o.})$, and Eq.~(\ref{eq:relation-Omega-qbarq}) must be replaced by
\begin{align}
\Omega_{\rm DM,0}h^2=2.8\times 10^8\bigg(2Y_{\chi^*}(x_{\rm f.o.})+\frac{122}{144}Y_{\Delta q,0} \bigg)\frac{m_\chi}{\rm GeV}\,.
\end{align} 
This scenario is sketched in the right diagram of Fig.~\ref{fig:boxdiag_final}. In this case, the initial asymmetry in the dark sector necessary to reproduce the quark-antiquark asymmetry is still given by Eq.~(\ref{eq:YDeltaQ-0}). However, since the total dark matter yield is larger, the observed dark matter abundance is reproduced for a smaller value of the dark matter mass,
\begin{align}
m_\chi\Big|_{Y_{\chi*}(x_{\rm f.o.})\neq 0}=\frac{Y_{\Delta \chi}(x_{\rm f.o.})}{Y_{\Delta \chi}(x_{\rm f.o.})+2Y_{\chi^*}(x_{\rm f.o.})}\,m_\chi\Big|_{Y_{\chi*}(x_{\rm f.o.})=0}.
\end{align}
Here $Y_{\chi^*}(x)$ can be calculated from particularizing Eq.~(\ref{eq:Yk}) to the weak wash-out regime,
\begin{align}
    \frac{dY_{\chi^\ast}}{dx}&= - \frac{\lambda}{x^2}   \0{\langle \sigma v \rangle _{\chi\chi^\ast\to N \overbar{N}}}\Big(Y_{\chi^\ast}(Y_{\chi^\ast}+Y_{\Delta \chi}^{\rm in}) - 9\, P \,Y_{\chi}^{\rm eq, in} \Big).
    \label{eq:BE-Y*-weak}
\end{align}
Since at freeze-out the dark matter antiparticles were still in thermal equilibrium, one can approximate $Y_{\chi^*}(x_{\rm f.o.})=Y^{\rm eq}_{\chi^*}(x_{\rm f.o.})$, and then the equilibrium distribution can be easily obtained by setting the right hand side of Eq.~(\ref{eq:BE-Y*-weak}) to zero. We obtain
\begin{align}
 Y^{\rm eq}_{\chi^*}(x) \simeq \frac{-Y^{\rm in}_{\Delta\chi}}{2}+\sqrt{\frac{Y_{\Delta \chi}^{\rm in\,2}}{4}+9P(x)\left(Y_\chi^{\text{eq,in}}\right)^2}\,.
\label{eq:ChiEqUs}
\end{align}

Let us stress that these conclusions are quite insensitive to the concrete values of the portal strengths $\Lambda_2$, $\Lambda_0$ and $\Lambda_n$, provided that (i) the washout of the asymmetry is weak, (ii) $\Lambda_n$ is small enough to keep the hidden sector thermalized with the visible sector and (iii) $\Lambda_0\ll \Lambda_n$, so that $N$ is stable in the timescale of the freeze-out. Other scenarios are also possible, by adjusting the initial conditions in the Boltzmann equations. On the other hand, a crucial assumption of our scenario is the existence of a neutron portal that transmits the asymmetry in the dark sector to the visible sector. This neutron portal could lead to experimental signatures in our scenario, which will be discussed in detail in the next section.

\section{Constraints on the neutron portal}\label{sec:constraintNP}

\begin{figure}[t]
\centering
\includegraphics[width=0.6\textwidth]{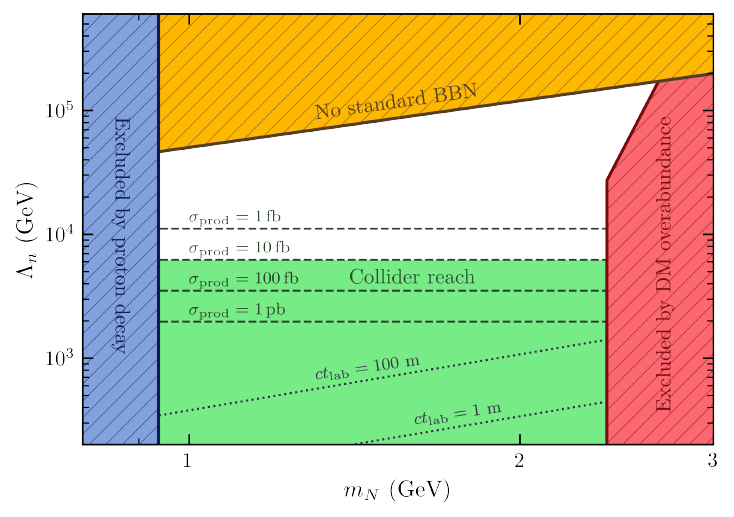}
\vspace{-0.3cm}
\caption{Constraints on the neutron portal energy scale ($\Lambda_n$) and mass of $N$ ($m_N$) from cosmology, proton stability, and collider experiments, along with contours of production cross section and decay length at the LHC with $\sqrt{s}=14$ TeV. The allowed region is shown in white.}
\label{fig:NeutronPortalConstraints}
\end{figure}

The particle $N$ can have implications in our visible sector through the neutron portal Eq.~(\ref{eq:LagNeutronPortal}), {\it e.g.} through the decay of $N$ into quarks, the production of $N$ in proton-proton collisions, or the generation of a mass mixing term with the neutron below the QCD confinement scale. The various constraints are summarized in   Fig.~\ref{fig:NeutronPortalConstraints}, in the parameter space defined by the mass of $N$ ($m_N$) and the energy scale of the neutron portal ($\Lambda_n$). The region allowed by all the constraints, shown in white, is bounded and could in principle be probed in its totality. Let us describe in detail the multiple constraints from the parameter space.

The standard Big Bang Nucleosynthesis (BBN) scenario is extremely successful in describing the evolution of the Universe after $\sim 1$ s. In particular, observations indicate that the quark-antiquark asymmetry at the time of BBN does not differ significantly from the quark-antiquark asymmetry at the time of recombination. In order to preserve the standard BBN scenario, we will require that the yield of $N$ is largely depleted at $\sim 1$ s, so that their decays have practically no impact at later times. Using the fact that the width of $N$ is given by
\begin{align}\label{eq:NPartonicLifetime}
\Gamma^{-1}_{N\rightarrow udd} \approx 1.6 \text{ s}\left(\frac{\n{\Lambda_n}}{10^5 \text{ GeV}}\right)^4\left(\frac{\text{GeV}}{m_N}\right)^5,
\end{align}
and requiring conservatively $\Gamma^{-1}_{N\rightarrow udd}\lesssim 0.1$ s, we exclude the region $\Lambda_n\gtrsim 10^5\,{\rm GeV} (m_N/{\rm GeV})^{5/4}$, indicated in Fig.~\ref{fig:NeutronPortalConstraints} as a hatched orange region.

The neutron portal also leads to the production of $N$ in proton-proton collisions through the partonic processes $u d\rightarrow N\bar d$ and  $d d\rightarrow N\bar u$. We estimate the non-resonant production cross section at $\sqrt{s}=14$ TeV to be
\begin{align}
\sigma_{p p \rightarrow N +\text{jet}} \approx 2\, {\rm fb} \left(\frac{f_{\rm PDF}}{10^{-2}}\right)
\left(\frac{10^4\,{\rm GeV}}{\Lambda_n}\right)^4\;,
\end{align}
where we have estimated the effect of the partonic distributions in the protons in the parameter to be $f_{\rm PDF}\approx 10^{-2}$ \cite{Clark:2016jgm}. Depending on the lifetime of $N$, the signal at colliders could be in the form of missing $p_T$ (if stable within the detector's volume), in the form of a displaced vertex (if the decay length is macroscopic), or in the form of dijets (if the decay length is microscopic). We show in Fig. \ref{fig:NeutronPortalConstraints} the line for which the decay length lies outside of the ATLAS or CMS detectors, $c t_{\rm lab}=100$ m, where we have taken a Lorentz factor $\gamma=\sqrt{\hat s}/(2m_N)$ with $\sqrt{\hat s}\sim 2$ TeV for the partonic center of mass energy. We also indicate in the plot the values of $\Lambda_n$ corresponding to a production cross section $\sigma_{p p \rightarrow N+ \text{jet}}=1$ fb, 10 fb, 100 fb  and 1 pb, and in green the ballpark area of values that can be probed at the LHC with an integrated luminosity of  ${\cal L}=100$ fb$^{-1}$, which corresponds to effective interactions with strength $\Lambda_n\lesssim 6$ TeV. We note that for small values of $\Lambda_n$ the effective field theory breaks down at LHC energies, and instead a dedicated search for the new particles mediating the neutron portal should be performed. A detailed collider analysis is however beyond the scope of this paper. We also note that the collider constraints become weaker if the portal between the dark sector and the visible sector involves sea quarks, {\it e.g.} $\overbar N s_R \overline{c_R^c} s_R$ instead of  $\overbar N d_R \overline{u_R^c} d_R$; this variant of the portal could be probed in the decays of heavy mesons and baryons \cite{Heeck:2020nbq}. 

The mass of $N$ is bounded from below from the requirement that the proton must be the lightest fermion carrying baryon number. Otherwise the proton could decay, {\it e.g.} $p\rightarrow N \pi^+$. This requirement translates into the lower limit $m_N \geq 938$ MeV (this limit on the mass could be avoided if the  width is suppressed by a large $\Lambda_n$, however this region is in tension with BBN \cite{McKeen:2020oyr}). Lastly, the mass of $N$ is bounded from above from the requirement that the dark matter is not overproduced. More specifically, the annihilation process $\chi\chi^*\rightarrow N \overbar{N}$ must be efficient enough to deplete most of the dark matter density. Naively, this requires $m_N< m_\chi$, however, as argued in \cite{Griest:1990kh,DAgnolo:2015ujb}, the annihilation can also occur in a ``forbidden'' channel, due to the existence of sufficiently energetic dark matter particles in the tail of the Maxwell-Boltzmann distribution. The thermally averaged annihilation cross section in a forbidden channel is approximately given by~\cite{DAgnolo:2015ujb}
\begin{align}\label{eq:ForbiddenAnn}
\0{\left\langle\sigma v\right\rangle}_{\0{\chi \chi^\ast \rightarrow N \overbar{N}}}\approx 8\pi f(\Delta) \0{\left\langle\sigma v\right\rangle}_{\0{ N \overbar{N} \rightarrow\chi \chi^\ast}}e^{-2\Delta x},
\end{align}
where $\Delta=\left(m_N-m_\chi\right)/m_\chi$ is the relative mass splitting and $f(\Delta)$ is a function of $\Delta$, which approximates to $f(\Delta) \approx 1+\Delta$ for $\Delta\gg1$. In Eq.~(\ref{eq:ForbiddenAnn}) the Boltzmann suppression of the dark matter annihilation cross section is explicit. 
A more rigorous upper limit on $m_N$ is derived by ensuring that dark matter overproduction is avoided for the largest possible value of the annihilation cross section, Eq.~(\ref{eq:ForbiddenAnn}). From the s-wave unitarity requirement $\0{\left\langle\sigma v\right\rangle}_{\0{N \overbar{N} \rightarrow \chi \chi^\ast}}\leq 4\pi/m_N^2 (x_{\text{f.o.}}/\pi)^{1/2}$, we obtain $m_N\lesssim 2.7$ GeV. For the milder requirement that the effective field theory remains valid, which corresponds to $\0{\Lambda_0}\sim m_\chi$, we obtain $m_N\lesssim 2.4$ GeV. This limit is shown in Fig. \ref{fig:NeutronPortalConstraints} in red, where the relaxation of constraint at large $\Lambda_n$ values is attributed to the freeze-out of dark matter taking place during a period of matter domination. In this case, the value of the total annihilation cross section required to obtain the correct dark matter relic abundance is smaller than in the standard WIMP paradigm \cite{Hamdan:2017psw}.

\section{Constraints from Dark Matter searches}\label{sec:DMconstr}
\subsection{Dark matter signals via the Higgs portal}

The Higgs portal $\lambda_{\chi H} |\chi|^2|H|^2$ leads to potential dark matter signals in collider experiments and in direct detection experiments (signals in indirect detection experiments could also arise if there is a relic population of dark matter antiparticles). In our scenario, the predicted dark matter mass is $\simeq 1.9\text{ GeV}$; therefore, the Higgs portal could induce the invisible decay of the Higgs into a dark matter particle-antiparticle pair, $h\rightarrow \chi \chi^{\ast}$.   The rate reads
\begin{align}
\Gamma_{h\rightarrow \chi\chi^*}\simeq \frac{\lambda_{\chi H}^2 v^2}{32\pi m_h}\;,
\end{align}
where $v\simeq 246$ GeV is the Higgs vacuum expectation value and $m_h\simeq 125$ GeV is the Higgs mass. Current experimental searches constraint the Higgs branching ratio into invisible final states to be $\text{Br}\left(h\rightarrow \text{ invisible}\right)\lesssim 20\%$ \cite{ATLAS:2015ciy, ATLAS:2022yvh}. Using that the Higgs width into visible particles is $\Gamma_{\rm vis}\simeq 4$ MeV, one obtains  $\lambda_{\chi H}\lesssim 10^{-2}$ \cite{Arcadi:2019lka}.

The Higgs portal interaction also induces the scattering of dark matter particles off nuclei. The spin-independent scattering cross section off a nucleon ${\cal N}$ reads~\cite{Cline:2013gha,McDonald:1993ex}
\begin{align}
    \sigma_{\chi {\cal N} \to \chi {\cal N}}&\simeq \frac{\lambda^2_{\chi H} f_{\cal N}^2}{\pi} \frac{\mu^2 m_{\cal N}^2}{m_h^4 m_\chi^2},
\end{align}
where $\mu=m_{\cal N}\, m_\chi/(m_{\cal N}+m_\chi)$ is the dark matter-nucleon reduced mass, and $f_{\cal N}\approx 0.3$ encodes the quark and gluon content of a nucleon. For dark matter in the GeV mass range, the best current sensitivity is provided by the DarkSide 50 experiment~\cite{DarkSide-50:2022qzh}, which excludes cross sections larger than $2\times 10^{-42}\text{ cm}^2$ for a dark matter mass of 1.9 GeV. This limit translates into $\lambda_{\chi H}\lesssim 0.07$, less sensitive than the constraint from the invisible Higgs decay.

In addition to these phenomenological constraints on the Higgs portal coupling, it is worthwhile mentioning a theoretical constraint stemming from the naturalness of the dark matter mass. The dark matter mass term receives after electroweak symmetry breaking a contribution $\delta m_\chi^2=\lambda_{\chi H}v^2$. Therefore, our favored value $m_\chi\simeq 1.9$ GeV points to $\lambda_{\chi H}<6\times 10^{-5}$, unless there is a fine cancellation between the mass term in the Lagrangian Eq.~(\ref{eq:LagKin}) and the contribution to the mass from the electroweak symmetry breaking. This strong constraint would make any Higgs portal-induced dark matter signal very difficult to detect.

\subsection{Dark matter signals via the neutron portal}

\begin{figure}[t]
\begin{center}
\begin{tikzpicture}
 \begin{feynman}[every dot={/tikz/fill=blue!50,/tikz/inner sep=2pt}]
    \vertex [dot](vm){};
    \vertex [above=1.2cm of vm] (va2);
    \vertex [left=1.2cm of va2] (va1){\(\chi\)};
    \vertex [right=1.2cm of va2] (va3){\(\chi\)};
    \vertex [above=0.4cm of vm] {\(\frac{1}{\Lambda_0}\)};
   \vertex [below=1.2cm of vm] (vc2);
   \vertex [left=0.55cm of vc2, crossed dot] (vc1){};
   \vertex [right=0.55cm of vc2, crossed dot] (vc3){};
            \vertex [left=0.45cm of vc1] (vc1dm){\(\)};
    \vertex [above=0.25cm of vc1dm] (vc1dm2){\(\delta m\)};  
   \vertex [right=0.45cm of vc3] (vc3dm){\(\)};
       \vertex [above=0.25cm of vc3dm] (vc1dm2){\(\delta m\)};  
   \vertex [below=2cm of vm] (vb2);
   \vertex [left=1.2cm of vb2] (vb1);
   \vertex [right=1.2cm of vb2] (vb3);    
   \vertex [below=0.2cm of vb1] (cap1);
   \vertex [below=0.2cm of vb3] (cap2);
    
    \vertex [below=2.2 cm of vm] (vbb2);
    \vertex [left=1.6cm of vbb2] (vbb1){\({\rm Xe}\)};
    \vertex [right=1.6cm of vbb2] (vbb3){\({\rm Xe}\)};
    
    \vertex [above=0.2cm of vbb2] (vbba2);
    \vertex [left=1.55cm of vbba2] (vbba1);
    \vertex [right=1.55cm of vbba2] (vbba3);
            
   \vertex [below=0.4cm of vb1] (cap1);
   \vertex [below=0.4cm of vb3] (cap2);

   \diagram* {
    (vm) -- [fermion](va3),
    (va1) -- [fermion] (vm),
    (vb1) -- [fermion, edge label=\(n\)] (vc1),
    (vc1) -- [fermion, edge label=\(N\)] (vm),
	(vc3) -- [fermion, edge label=\(n\)] (vb3),
	(vm) -- [fermion, edge label=\(N\)] (vc3),
	(vbba3) -- [plain] (vb3),	
    (vb1) -- [plain] (vbba1),
    (cap1) -- [plain, white, edge label'=\(\color{black}{\rm a)}\)] (cap2)
    };
 \end{feynman}
 \draw [line width=0.5mm] (vbb1) -- (vbb3);
\end{tikzpicture}\qquad\qquad\begin{tikzpicture}
\begin{feynman}[every dot={/tikz/fill=purple!50,/tikz/inner sep=2pt}]
\vertex [dot](vm){};
   \vertex [above=1.2cm of vm] (va2);
   \vertex [left=1.2cm of va2] (va1){\(\chi\)};
   \vertex [right=1.2cm of va2] (va3){\(\chi^\ast\)};
   \vertex [above=0.4cm of vm] {\(\frac{1}{\Lambda_2}\)};
   \vertex [below=1cm of vm] (vc2);
  \vertex [left=0.55cm of vc2, crossed dot] (vc1){};
   \vertex [right=0.55cm of vc2, crossed dot] (vc3){};
            \vertex [left=0.45cm of vc1] (vc1dm){\(\)};
    \vertex [above=0.25cm of vc1dm] (vc1dm2){\(\delta m\)};  
   \vertex [right=0.45cm of vc3] (vc3dm){\(\)};
       \vertex [above=0.25cm of vc3dm] (vc1dm2){\(\delta m\)};  
   \vertex [below=2cm of vm] (vb2);
   \vertex [left=1.2cm of vb2] (vb1){\(n\)};
   \vertex [right=1.2cm of vb2] (vb3){\(\overbar{n}\)};    
   \vertex [below=0.4cm of vb1] (cap1);
   \vertex [below=0.4cm of vb3] (cap2);
   
  \diagram* {
   (va3) -- [fermion](vm),
   (va1) -- [fermion] (vm),
   (vc1) -- [fermion, edge label=\(N\)] (vm),
   (vb1) -- [fermion] (vc1),
	(vb3) -- [fermion] (vc3),
	(vc3) -- [fermion, edge label'=\(\overbar{N}\)] (vm),
    (cap1) -- [plain, white, edge label'=\(\color{black}{\rm b)}\)] (cap2)
    };
    \end{feynman}
\end{tikzpicture}

\end{center}
\vspace{-0.5cm}
\caption{Processes induced by the neutron portal: dark matter scattering off nuclei (left panel) and dark matter-induced ``transmutation" of a neutron into an antineutron (right panel).
}\label{fig:DDDiag}
\end{figure}
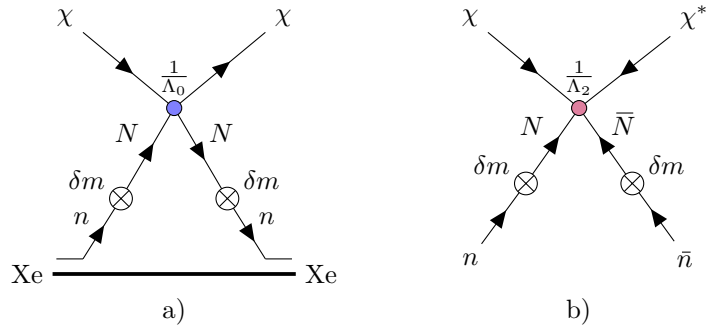

The second portal of the dark sector to the Standard Model is the neutron portal $\frac{1}{\Lambda_n^2}\Nbar d_R \, \overline{u_R^c} d_R$. This term induces, at energies below the QCD confinement scale, a mass mixing term between the dark matter particles and the neutrons
\begin{align}
\delta m \simeq 1.4 \times 10^{-8}\left(\frac{ 10^3 \text{ GeV}}{\n{\Lambda_n}}\right)^2\text{ GeV}\;,
\end{align}
which leads to the scattering of dark matter particles with nuclei and the ``transmutation'' of neutrons into antineutrons, as depicted in Fig.~\ref{fig:DDDiag}. 

We can estimate the rate of dark matter scatterings off neutrons as
\begin{align}
    \sigma_{\chi n \to \chi n}&\simeq\frac{1}{4\pi}\left(\frac{\delta m}{m_N}\right)^4 \frac{ m_n^2 }{(m_n + m_\chi)^ 2} \frac{1}{\Lambda_0^2} \nonumber\\
    &\simeq 10^{-61} \text{~cm}^2\left(\frac{{\rm GeV}}{\Lambda_0}\right)^2\left(\frac{10^3\,{\rm GeV}}{\Lambda_n}\right)^8,~\label{eq:sigmachin}
\end{align}
which is many orders of magnitude below the current sensitivity of the DarkSide 50 experiment~\cite{DarkSide-50:2022qzh} and of any foreseeable dark matter direct detection experiment. We also do not expect to have detectable signals of dark matter scatterings from neutron stars.

In principle, the ``transmutation" of neutrons into antineutrons $\chi n \rightarrow \chi^\ast \overbar{n}$ could also lead to observable signatures.\footnote{Previous works have considered this effect in the context where $N$ is the dark matter candidate \cite{Jin:2018moh, Fornal:2020poq}. Other scenarios that link dark matter and baryogenesis can also generate similar exotic decays of baryons \cite{Berger:2023ccd}.} We estimate the cross section for this process to be
\begin{align}
    \sigma_{\chi n \to \chi^* \bar n}\simeq 10^{-83} \text{~cm}^2\left(\frac{10^{11}\,{\rm GeV}}{\Lambda_2}\right)^2\left(\frac{10^3\,{\rm GeV}}{\Lambda_n}\right)^8.~\label{eq:sigmachintransmutation}
\end{align}
which we obtain by simply replacing $\Lambda_0$ by $\Lambda_2$ in Eq.~(\ref{eq:sigmachin}). We have also normalized the ``transmutation" cross section to $\Lambda_2=10^{11}$ GeV, which is a typical value for which the particle-antiparticle asymmetries are not significantly washed out, {\it cf.} Eq.~(\ref{eq:L2wo}). The searches for neutron-antineutron oscillation at Super-Kamiokande~\cite{Super-Kamiokande:2020bov} can be recast into a limit on the neutron-antineutron transmutation rate induced by dark matter particles in the Milky Way halo. We estimate this limit to be $\sigma_{\chi n \to \chi^* \bar n}\lesssim 10^{-49}\,{\rm cm}^2$, which is far above the expected cross section in our framework.

The prospects for indirect dark matter detection are strongly influenced by whether the symmetric component has been depleted at freeze-out. In the asymmetric case, the only annihilation channel available is $\chi \chi \rightarrow \overbar{N}\overbar{N}$. This process is highly suppressed in the present realization due to the large effective scale $\2{\Lambda_2}$, and no constraints on the model parameters are expected. Nevertheless, it is noteworthy that an asymmetric complex scalar dark matter candidate could have an open direct annihilation channel today. In the symmetric scenario, it is expected that strong $\chi \chi^\ast \rightarrow N \overbar{N}$ annihilations will occur, as $\chi$ is a light and thermal dark matter candidate. However, due to the conservation of baryon number, $N$ must eventually decay into neutrons, whose masses would already account for a significant portion of the energy budget in the final state of the annihilation. The decay of $N$ into semi-relativistic neutrons would emit a soft cascade of various hadrons and/or a gamma-ray of energy $E_\gamma=m_N-m_n$, which could provide a window into both the nature of dark matter and its link to the neutron portal.

\section{Conclusions}
\label{sec:conclusions}

We have presented a scenario that accommodates both dark matter and a  quark-antiquark asymmetry.
We have postulated the existence of a spin-0 and a spin-1/2 particle in the dark sector, both singlets under the Standard Model gauge group and carrying baryon number. As initial conditions, we have assumed that at very high temperatures, the Universe has zero baryon and lepton numbers but that the dark sector contains an asymmetry between particles and antiparticles. We have argued that the asymmetry in the spin 1/2 particle can be transmitted to the visible sector through (baryon conserving) neutron portal interactions, thus resulting in a quark-antiquark asymmetry (and possibly a lepton-antilepton asymmetry via sphalerons). On the other hand, the spin-0 particle is stable due to the baryon number conservation and constitutes a dark matter candidate. In this framework, the $B-L$ number of the visible sector is exactly compensated by an opposite asymmetry in the dark sector, thus linking the observed quark-antiquark asymmetry to the existence of dark matter.

Under reasonable assumptions, we expect the dark matter mass to be $\sim 1.9$ GeV if it is fully asymmetric or potentially lighter if a population of dark matter antiparticles remains after freeze-out. The scenario also predicts the mass of the exotic spin-1/2 particle to be comparable to that of the dark matter. Such a particle could be produced at the LHC or in flavor physics experiments through the neutron portal, generically leaving the detector before decaying. This particle would then produce a signal of missing energy and an apparent violation of baryon number due to the imbalance in the baryon number of the visible sector particles involved in the reaction. 

Finally, we have briefly discussed the prospects for observing dark matter signals in our scenario. We find that the most promising avenue to detect signals lies in the Higgs portal, either through the invisible Higgs decay width or in direct detection experiments, akin to the singlet scalar dark matter model. 

\acknowledgments{
\noindent This work is supported by the Collaborative Research Center SFB1258 and by the Deutsche Forschungsgemeinschaft (DFG, German Research Foundation) under Germany's Excellence Strategy - EXC-2094 - 390783311. We thank Motoko Fujiwara, Florian Jörg, Robert McGehee and David Rousso for useful discussions. We thank in particular Nick Proff for discussions and pointing out a mistake in the asymmetry dynamics.}

\normalem
\bibliographystyle{JHEP}
\bibliography{biblio}

\end{document}